\documentclass[a4paper,12pt]{article}

\usepackage[T2A]{fontenc}
\usepackage{amsmath,amssymb}
\usepackage{cite}
\usepackage[unicode,linktocpage=true,plainpages=false,pdfpagelabels=false]{hyperref}

\newcommand{\dd}{\partial}
\newcommand{\de}{\delta}
\newcommand{\De}{\Delta}
\newcommand{\m}{\mu}
\newcommand{\n}{\nu}
\newcommand{\ls}{\left(}
\newcommand{\rs}{\right)}
\newcommand{\La}{\Lambda}
\newcommand{\ka}{\varkappa}
\newcommand{\dz}{\zeta}
\newcommand{\ga}{\gamma}
\newcommand{\ff}{\varphi}
\newcommand{\ta}{\tau}
\newcommand{\al}{\alpha}
\newcommand{\be}{\beta}
\newcommand{\Ga}{\Gamma}
\newcommand{\ps}{\psi}
\newcommand{\np}{\emptyset}
\newcommand{\tri}[1]{\overset{{\scriptscriptstyle 3}}{#1}{}}
\newcommand{\om}{\omega}

\newcommand{\disn}[2]{$$\displaylines{\refstepcounter{equation}%
            \label{#1}\hskip 1em minus 1em #2\hfilneg}$$}
\newcommand{\nom}{\hfil\hskip 1em minus 1em (\theequation)}
\newcommand{\no}{\hfil \hskip 1em minus 1em\phantom{(\theequation)}%
            \hfilneg\cr\hfilneg\hskip 1em minus 1em\hfil}

\begin{document}

\title{Non-relativistic limit of embedding gravity\\ as General Relativity with dark matter}

\author{S.~A.~Paston\thanks{E-mail: pastonsergey@gmail.com}\\
{\it Saint Petersburg State University, Saint Petersburg, Russia}
}
\date{\vskip 15mm}
\maketitle

\begin{abstract}
Regge-Teitelboim embedding gravity is the modified gravity based on a simple string-inspired geometrical principle: our spacetime is considered here as a 4-dimensional surface in a flat bulk.
This theory is similar to the recently popular theory of mimetic gravity: the modification of gravity appears in both theories as a result of the change of variables in the action of
General Relativity. Embedding gravity, as well as mimetic gravity, can be used in explaining the dark matter mystery since, in both cases, the modified theory can be presented as General Relativity with additional fictitious matter (embedding matter or mimetic matter).
For the general case, we obtain the equations of motion of embedding matter in terms of embedding function as a set of first-order dynamical equations and constraints consistent with them. Then we construct a non-relativistic limit of these equations, in which the motion of embedding matter turns out to be slow enough so that it can play the role of cold dark matter.
The non-relativistic embedding matter turns out to have a certain self-interaction, which could be useful in the context of solving the core-cusp problem that
appears in the $\La$CDM model.
\end{abstract}

\newpage

\section{Introduction}\label{vved}
In the attempts to explain some astronomical observations in the framework of General Relativity (GR), certain discrepancies arise, which can be solved by an assumption of the existence of an additional matter. This matter is called Dark Matter (DM) since it can not be directly observed. Such discrepancies appear at a wide range of scales: from galactic to Hubble ones, see, e.g.,  \cite{1611.09846}. In the case of galactic, it is the deviation of the stars' motion from the GR predictions in the assumption that stars are moving only in the gravitational field of visible matter ("rotation curves"{}). In order to explain this discrepancy, an assumption is made that there is some amount of DM in galaxies besides ordinary matter. The density of DM decreases slower than ordinary matter density, so it forms a halo around a galaxy. At the larger scales, the observations of gravitational lensing, interpreted in the GR framework, indicate a significant amount of matter in the regions with a low amount of visible matter. This fact can also be explained using the assumption of the presence of DM in such regions. At the largest cosmological scale, the assumption of the homogeneous distribution of DM (in the amount of 26\% of the Universe total {mass-energy}), alongside with the existence of dark energy (69\%) whose properties are close to the cosmological constant, ensures the matching of average {mass-energy} density with the critical density, as well as provides the solution of the structure formation problem.
For a historical survey of the DM-related observational discoveries and theoretic arguments, see the review \cite{1605.04909}.

The majority of the currently
available observational data (including the very important data related to the cosmic microwave background anisotropy) can be satisfactorily described by $\La$-Cold Dark Matter ($\La$CDM) model, which is a Standard Model of modern cosmology. In its framework, DM can be considered as non-relativistic dust matter, which produces the same gravitational field as ordinary matter, but
non-gravitational interaction of dark matter with ordinary one is either absent or very weak and thus undetectable \cite{2005.03520}.
It should, however, be noted that some observations
on the scales of galaxies remain not quite consistent with
N-body simulations in $\La$CDM model.
As examples of such inconsistencies, we can mention
the \emph{core-cusp} problem, as well as the \emph{missing satellites}
and \emph{too-big-to-fail} problems, see details in recent reviews \cite{1606.07790,1707.04256,2007.15539}.

Numerous attempts to detect DM have been fruitless so far \cite{1509.08767,1604.00014}, which makes a mystery of the DM nature one of the most intriguing problems of modern physics. There are many hypotheses (see, e.g., \cite{1703.07364,astro-ph/0003365,1705.02358}) related to the nature of DM, the choice of which determines a way to detect it. The failures of detecting can be connected to the weakness of the coupling between the dark and the ordinary matter. However, a more radical explanation exists. Since DM undoubtedly interacts with gravity, the absence of its participation in other interactions can be interpreted as an indication of the fact that DM is not a \emph{real} matter coupled with {Einsteinian} gravity, but rather is an effect of a description of the gravitational interaction in the framework of a certain \emph{modified} theory of gravity. In such an approach, DM turns out to be absent from the point of view of the \emph{fundamental} theory,
but it reveals its existence
through interaction with gravity (via a contribution to Einstein equations) if the latter is described in the framework of GR instead of its description in the context of the modified theory.

One possible candidate for the role of this modified theory is MOND \cite{mond},
for the review of attempts to explain observations in the MOND framework see, e.g., \cite{2007.15539} and the references therein.
However, modifications that have additional degrees of freedom in comparison with GR seem more promising (see the review of modifications in \cite{1108.6266}, see also \cite{1708.00603}
and \cite{2007.15539}).
In consideration of modified theory in the context of GR, these degrees of freedom turn out to correspond to DM.
In recent years strong attention has been drawn to the mimetic gravity \cite{mukhanov}, which allows one to introduce "dark matter without dark matter"{} \cite{mukhanov} in the described manner. This model appears as a result of the change of variables, which isolates the conformal mode of the metric. Since this change of variables involves differentiation (being a particular case of the differential field transformations in GR \cite{statja60}), it leads to the extension of dynamics of the theory, and some additional degrees of freedom w.r.t. GR appear. These degrees of freedom describe the \emph{mimetic matter}, which might be interpreted as DM. The possibility to reformulate the mimetic gravity as the GR with an additional contribution to the action \cite{Golovnev201439} shows that the mimetic matter is a potentially moving dust effectively described by two scalar fields. The restriction of the potentiality of motion can be lifted \cite{statja48}, the introduction of a pressure is also possible \cite{lim1003.5751}, but the theory nevertheless requires significant extensions to resolve the discrepancies mentioned above in the framework of mimetic gravity, see, e.g., \cite{1601.00102} and also review \cite{mimetic-review17}.

Another example of modified theory, whose origins lie in the procedure of differential field transformations in the GR, is Regge-Teitelboim embedding gravity (also known as \emph{embedding theory}) \cite{regge}. It appears as a result of the following change of variables:
 \disn{r1}{
g_{\m\n}=(\dd_\m y^a)(\dd_\n y^b)\,\eta_{ab}
\nom}
(here and hereafter $\m,\n,\ldots=0,1,2,3$)
in the Einstein-Hilbert action of GR.  This change of variables, in contrast with the mimetic one, has a deep geometric nature, since \eqref{r1} is a formula for induced metric on the 4-dimensional surface described by embedding function $y^a(x^\m)$ in an ambient spacetime with Minkowski metric $\eta_{ab}$. Therefore the embedding gravity is based on a simple string-inspired assumption that our spacetime is not merely an abstract pseudo-Riemannian space but rather a surface in a flat ambient spacetime of higher dimension.
It should be stressed that in the present approach extra dimensions appear in the \emph{ambient} space, whereas our spacetime remains four-dimensional, so the compactification problem, which is well known in the string theory,  does not arise here.
In accordance with the Janet-Cartan-Friedman theorem
\cite{fridman61}, the minimal number of dimensions of a spacetime in which an arbitrary 4-dimensional surface can be \emph{locally} embedded, is equal to ten, so the ambient spacetime in embedding gravity is usually taken to be 10-dimensional. In the present paper we will assume that as well, so $a,b,\ldots=\np,1,\ldots,9$.

After the appearance of paper \cite{regge} the ideas of embedding gravity were discussed in the paper \cite{deser} and were used lately by many authors with the aim of description of gravity, including its quantization (see, e.g., \cite{pavsic85,tapia,estabrook1999,davkar,statja18,rojas09,statja25,faddeev,statja26,statja33}).
In this framework, one can try to explain the effects of DM in the assumption of a strict spatial homogeneity and isotropy of the universe  \cite{davids01,statja26}.
To do it at the smaller scales, one might try to reformulate the theory as the GR with an additional contribution corresponding to some fictitious matter \cite{statja51} and to examine the properties of this \emph{embedding matter}. The number of variables that describe embedding matter turns out to be sufficiently large, and the properties of this matter are sufficiently rich to hope for a successful description of observational properties of DM in the framework of embedding gravity, without the introduction of additional extensions (as it is done, e.g., in the framework of mimetic gravity).

The fact that embedding gravity is based on the simple geometric idea without any \emph{ad hoc} extensions and arbitrary constants , as well as the potential quantization advantages connected with the presence of flat ambient spacetime in the formulation of the theory (see, e.g., \cite{pavsic85,davkar}), makes this modification of gravity, in our opinion, an especially promising in the search for an explanation of the mystery of DM.

In the present paper, we will analyze the equations of motion of embedding matter in the \emph{non-relativistic limit}, whose definition was proposed in the paper \cite{statja67}. In this analysis, we will follow the paper \cite{statja67} in part, but the calculations will be performed in a more straightforward and geometrically clear manner. In the \cite{statja67} a non-square vielbein $e^a_\m\equiv\dd_\m y^a$ was used instead of embedding function $y^a$ (such an approach was also used in another variant of embedding gravity \cite{faddeev}), and such vielbein was required to satisfy the integrability condition $\dd_\m e_\n^a=\dd_\n e_\m^a$. Due to this choice of independent variables, the embedding function $y^a$, which describes 4-dimensional surface, appeared in the calculations only in an implicit way. In the present paper, we show a straightforward way to obtain the same result, using the embedding function $y^a$ as one of the variables describing embedding matter.

In Section~\ref{emb}, we remind the formulation of embedding gravity, both in its original form, based on the differential field transformations in GR, as well as in the form of GR with additional embedding matter contribution in the action, and write down its equations of motion.
In Section~\ref{ur-dv}, we analyze and rewrite these equations in the form of dynamical equations (with first-order time derivatives only) for the independent variables describing embedding matter and constraints that restrict the choice of these variables at the initial moment of time.
In Section ~\ref{nonrel1}, we consider the equations of motion in the limiting case of non-relativistic approximation, whereas in Section~\ref{nonrel2} the deviations from this limit are considered.

We succeeded in showing that the embedding matter in the studied limit turns out to be a non-relativistic matter with certain self-interaction. If this self-interaction is not very strong, this behavior quite well corresponds to the known properties of DM at all scales, but
the very presence of self-interaction could be useful. For example, the models in which DM is the Self-Interacting Dark Matter (SIDM) \cite{1705.02358} can be considered as solutions to
 the well-known core-cusp problem. This problem manifest itself in the fact that according to calculations the density of DM without self-interaction must increase rapidly towards the center of a galaxy, forming "cuspy"{} profiles. However, the observations do not confirm this, giving "cored"{} profiles.
The further examination of the equations of embedding matter motion, obtained in the non-relativistic limit, would allow to determine its behavior in different regimes and to solve the question of applicability of the discussed approach to the solution of the core-cusp problem and other observational discrepancies on the scales of galaxies, some of which were mentioned above.

\section{Embedding gravity}\label{emb}
Regge-Teitelboim embedding gravity (or embedding theory) is the theory of modified gravity appearing as a result of the change of variables \eqref{r1} in the action of GR:
 \disn{n1}{
S=-\frac{1}{2\ka}\int\! d^4x\sqrt{-g}\,R+S_{\text{m}},
\nom}
where $S_{\text{m}}$ is a contribution of an ordinary matter. In this approach the independent variable describing gravity is an embedding function $y^a(x^\m)$
(we remind that $\m,\n,\ldots=0,1,2,3$; $a,b,\ldots=\np,1,\ldots,9$; we also assume that the signature of the flat ambient spacetime is $(+-\ldots-)$, so the metric $g_{\m\n}$ has the signature $(+---)$).
The variation of \eqref{n1} w.r.t. independent variable $y^a$ leads to the Regge-Teitelboim (RT) equations \cite{regge}
 \disn{r2}{
D_\m\Bigl(( G^{\m\n}-\ka\, T^{\m\n}) \dd_\n y^a\Bigr)=0,
\nom}
where $D_\m$ is a covariant derivative, $G^{\m\n}$ is the Einstein tensor and $T^{\m\n}$ is the EMT of the ordinary matter.

In order to simplify the analysis of the embedding gravity from the familiar general-rela\-ti\-vis\-tic point of view, one can rewrite the RT equations \eqref{r2} as a set of equations \cite{pavsic85}
 \disn{r3.1}{
G^{\m\n}=\ka \ls T^{\m\n}+\ta^{\m\n}\rs,
\nom}\vskip -2em
 \disn{r3.2}{
D_\m\Bigl(\ta^{\m\n}\dd_\n y^a\Bigr)=0.
\nom}
The first one can be interpreted as the Einstein equation with an additional contribution $\ta^{\m\n}$ to EMT from a certain \emph{embedding matter}, whereas the second one plays the role of the equation of motion for this fictitious matter. The equations  \eqref{r3.2} resembles a local conservation law for a set of some currents:
 \disn{s6.1}{
D_\m j^\m_a=0\quad\Leftrightarrow\quad
\dd_\m \ls\sqrt{-g}\,j^\m_a\rs=0,
\nom}\vskip -2em
 \disn{s6.2}{
j^\m_a=\ta^{\m\n}\dd_\n y_a.
\nom}

There is a possibility to reformulate the embedding gravity as GR with some additional matter not only at the level of equations of motion but also at the level of action.
To do this, one needs to add certain action of embedding matter $S^{\text{add}}$ to the action \eqref{n1}:
\disn{v2}{
S=S^{\text{EH}}+S_{\text{m}}+S^{\text{add}}.
\nom}
The metric here is assumed to be an independent variable, and the condition \eqref{r1} appears as a consequence of the equations of motion. In the paper \cite{statja51} several ways to choose $S^{\text{add}}$ were considered. Let us introduce two of them, which seem the most promising in the context of the analysis of embedding matter properties.

First of all, one can choose the symmetric tensor $\ta^{\m\n}$ alongside with embedding function $y^a$ as independent variables describing embedding matter \cite{statja48}:
\disn{za1}{
S_1^{\text{add}}=\frac{1}{2}\int\! d^4 x\, \sqrt{-g}\,
\Bigl( (\dd_\m y^a)(\dd_\n y_a) - g_{\m\n}\Bigr)\tau^{\m\n},
\nom}
so the independent quantity $\ta^{\m\n}$, which is playing the role of a Lagrange multiplier, turns out to be an EMT of embedding matter. As an alternative, one can assume the independence of the set of currents $j^\m_a$ in the equations \eqref{s6.1} instead of $\ta^{\m\n}$ ($y^a$ is still independent) \cite{statja51}:
\disn{r5}{
S_2^{\text{add}}=\int\! d^4 x\, \sqrt{-g}\,
\Bigl( j^\m_a\dd_\m y^a-\text{\bf tr}\sqrt{g_{\m\n}j^\n_a j^{\al a}}\Bigr).
\nom}
Here it is assumed that one needs to take the square root of matrix with the indices $\m$ and $\al$ and subsequently take the trace of the result (the operation \text{\bf tr}). In terms of $j^\m_a$ the EMT of embedding matter can be written as\disn{r5.1}{
\ta_\m{}^\al=\sqrt{g_{\m\n}j^\n_a j^{\al a}},
\nom}
where  the square root of matrix is again present, see details in \cite{statja51}.
Note that \eqref{r5.1} is consistent with \eqref{s6.2}.

The set of currents $j^\m_a$ can be interpreted as four vectors (indexed by $\m$) of the 10-dimensional flat bulk. In Sections~\ref{nonrel1} and \ref{nonrel2} we will analyze the behavior of embedding matter in such case that all these four vectors are \emph{non-relativistic in the bulk} \cite{statja67}, i.e.
 \disn{s7}{
j^\m_a=\de^\np_a j^\m+\de j^\m_a,\qquad
\de j^\m_a\to0.
\nom}
As we said in the Introduction, the analysis will be performed partially in spite of the paper \cite{statja67}, but with more straightforward and geometrically clear way of calculations.

If the expression \eqref{za1} is used as a contribution of embedding matter, the full set of its equations of motion has the form:
 \disn{s9.1}{
(\dd_\m y^a)(\dd_\n y^b)\,\eta_{ab}=g_{\m\n},
\nom}\vskip -2em
 \disn{s9.2}{
D_\m\Bigl(\ta^{\m\n}\dd_\n y^a\Bigr)=0,
\nom}
and the metric $g_{\m\n}$ needs to be found through a separately considered Einstein equation~\eqref{r3.1}
(the connection $\Ga^\al_{\m\n}$ that is present in the covariant derivative $D_\m$, is represented by Christoffel symbols which are expressed through the metric).
The form of equations of motion of embedding matter is different in the case where one takes the set of currents $j^\m_a$ as independent variables instead of $\ta^{\m\n}$, and \eqref{r5} is chosen as an action of the embedding matter, see \cite{statja51}.

\section{Equations of motion of embedding matter}\label{ur-dv}
Let us note that the satisfaction of \eqref{s9.1} leads to the relation that is well known in the formalism of the embedding theory (see, e.g., \cite{statja18}):
 \disn{s10}{
(\dd_\be y_a) D_\m \dd_\n y^a=0.
\nom}
It means that the second fundamental form of the surface
 \disn{n1.1}{
b^a_{\m\n}\equiv D_\m \dd_\n y^a
\nom}
is transverse w.r.t. index $a$, i.e. orthogonal to the surface. Using \eqref{s10}, one can rewrite \eqref{s9.2} as a set of two equations:
 \disn{s11.1}{
D_\m \ta^{\m\n}=0,
\nom}\vskip -2em
 \disn{s11.2}{
\ta^{\m\n}D_\m\dd_\n y^a=0.
\nom}

Let us consider the equations of motion of embedding matter \eqref{s9.1},\eqref{s11.1},\eqref{s11.2}
as a time $x^0$ dynamics of the independent variables $y^a$, $\ta^{\m\n}$.
Let us also recall another known formula of the theory of embedded surfaces, which connects the curvature tensor $R_{\al\be\m\n}$ and second fundamental form of the surface $b^a_{\m\n}$
(see, e.g., \cite{statja18}):
 \disn{n1.2}{
R_{\al\be\m\n}=b_{a\al\m} b^a_{\be\n}-b_{a\al\n} b^a_{\be\m}.
\nom}
Taking $\al=\m=0$, $\be=i$, $\n=k$ in it (here and hereafter $i,k,\ldots=1,2,3$) and using \eqref{n1.1},
we obtain the corollary of \eqref{s9.1}:
 \disn{n3}{
R_{0i0k}=(D_0 \dd_0 y_a) b^a_{ik} -(D_i \dd_0 y_a)(D_k \dd_0 y^a).
\nom}

The components of the second fundamental form of the surface $b_{ik}^a$ can be treated as a $6\times6$ matrix,
since the pair of indices $\{ik\}$, w.r.t. which this quantity is symmetric, have six possible values, and the index $a$ also can take only six values (due to the transversality condition \eqref{s10}) without the vanishing of the quantity.
In the generic case, where this matrix is invertible, one can use an "inverse"{} matrix, introducing the quantity
$\al^{ik}_a$, which is uniquely defined by the relations
 \disn{s16}{
\al^{ik}_a=\al^{ki}_a,\qquad
\al^{ik}_a \dd_\m y^a=0,\qquad
\al^{ik}_a b_{lm}^a=\frac{1}{2}\ls \de^i_l\de^k_m+\de^i_m\de^k_l\rs.
\nom}
Using this quantity, one can
{express}
the transverse part of $D_0\dd_0 y_a$ from \eqref{n3}, and since the {tangent} part of it vanished due to \eqref{s10}, one finds that
 \disn{n4}{
D_0\dd_0 y_a=\al^{ik}_a\big((D_i \dd_0 y_b)(D_k \dd_0 y^b)+R_{0i0k} \big).
\nom}

We used only one equation of motion in the derivation of the relation \eqref{n4}, namely \eqref{s9.1}.
Using \eqref{n4} in the another equations of motion \eqref{s11.2}, we obtain new equation
 \disn{n5}{
\ta^{00}\al^{ika}\big((D_i \dd_0 y_b)(D_k \dd_0 y^b)+R_{0i0k} \big)+2\ta^{0k}D_k\dd_0 y^a+\ta^{ik}D_i\dd_k y^a=0,
\nom}
which now can be used instead of \eqref{s11.2}.
Using the quantity $\al^{ik}_a$ again, one can extract the components of  $\ta^{ik}$ from this equation:
 \disn{s17}{
\ta^{lm}=-\al^{lm}_a \biggl( \ta^{00}\al^{ika}\Bigl((D_i \dd_0 y_b)(D_k \dd_0 y^b)+R_{0i0k}\Bigr) +2\ta^{0k}D_k \dd_0 y^a\biggr).
\nom}
We are therefore able to express the components of $\ta^{ik}$ through other variables, excluding it from the set of independent variables.

As a result, only $y^a$ and $\ta^{0\n}$ remain independent variables describing embedding matter, and the equations of motions of this matter are reduced to  \eqref{s9.1} and \eqref{s11.1} with \eqref{s17},
i.e. to the system of equations
 \disn{n6.11}{
(\dd_i y_a)(\dd_k y^a)=g_{ik},
\nom}\vskip -2em
 \disn{n6.12}{
(\dd_0 y_a)(\dd_k y^a)=g_{0k},
\nom}\vskip -2em
 \disn{n6.13}{
(\dd_0 y_a)(\dd_0 y^a)=g_{00},
\nom}\vskip -2em
 \disn{n6.21}{
D_0 \ta^{00}=-D_i \ta^{0i},
\nom}\vskip -2em
 \disn{n6.22}{
D_0 \ta^{0m}=D_l \Biggl(\al^{lm}_a \biggl( \ta^{00}\al^{ika}\Bigl((D_i \dd_0 y_b)(D_k \dd_0 y^b)+R_{0i0k}\Bigr) +2\ta^{0k}D_k \dd_0 y^a\biggr)\Biggr).
\nom}

The equation \eqref{n6.11}, in contrast with remaining equations \eqref{n6.12}-\eqref{n6.22}, does not contain time derivatives. In the canonical formulation (different forms of it for embedding theory was studied in the works \cite{regge,tapia,frtap,statja18,rojas09,statja24,statja35,rojas20})
such equations are called constraints. In the examination of constraints, after the discovery of primary constraints one usually needs to check their conservation in time in order to find secondary constraints. Let us find the conservation condition for the constraint
\eqref{n6.11}, using the covariant time derivative $D_0$ instead of ordinary one $\dd_0$ for the sake of convenience.
These are equivalent, since
{change from $\dd_0$ to $D_0$ leads only to the appearance of contribution} proportional to a constraint. We obtain:
 \disn{n6b}{
D_0 \big( (\dd_i y^a)(\dd_k y_a)\big)=(D_i \dd_0 y^a)(\dd_k y_a)+(\dd_i y^a)(D_k \dd_0 y_a),
\nom}
where we used that $D_0\dd_i y^a=D_i \dd_0 y^a$. Using the {product}
rule in \eqref{n6b} and taking \eqref{n6.12} into account, we obtain the condition of the conservation of constraint  \eqref{n6.11}:
 \disn{n6b1}{
(\dd_0 y_a) D_i \dd_k y^a=0
\nom}
(note that this equation is a particular case of the abovementioned relation \eqref{s10}).
Using the expression for a covariant derivative as well as relation  \eqref{n6.11},\eqref{n6.12},
one obtains
 \disn{n6a}{
(\dd_0 y_a) \dd_i \dd_k y^a=\Ga_{0,ik}.
\nom}
In the generic case, when the three vectors (at different values of $k$) $\dd_k y^a$
together with six vectors (at different values of symmetric indices $i,k$) $\dd_i \dd_k y^a$
form a basis of nine vectors of the ambient space, the equations  \eqref{n6.12} and \eqref{n6a} together with normalization equation \eqref{n6.13} allow to uniquely determine the quantity  $\dd_0 y_a$
if the function $y^a(x^0,x^i)$ at the given moment is known.

Such expression defines the dynamics of the embedding function  $y^a$, providing the satisfaction of  \eqref{n6.12} and \eqref{n6.13}, as well as the conservation of the constraint \eqref{n6.11} that must be imposed when one chooses $y^a$
at the initial moment of time $x^0$. The remaining equations \eqref{n6.21}, \eqref{n6.22} define the dynamics of the remaining independent variables $\ta^{00}$ and $\ta^{0m}$, so the equations of motion of embedding matter are reduced to the constraint \eqref{n6.11} and dynamical equations for independent variables
$y^a$ and $\ta^{0\n}$ which expresses their first time derivative through their values at the given moment of time.

In the general case dynamical equation for  $y^a$ turns out to be quite bulky, but it can nevertheless be written in an explicit form. To do that, let us notice that at the fixed  $x^0$ the embedding function  $y^a(x^0,x^i)$ defines a certain 3-dimensional surface in the ambient spacetime, so one can define a second fundamental form of it:
 \disn{nn1}{
\tri{b}^a_{ik}=\tri{D}_i\dd_k y^a,
\nom}
where $\tri{D}_i$ is a 3-dimensional covariant derivative,
which contains  3-dimensional Christoffel symbol $\tri{\Ga}^m_{ik}$ dependent on
3-metric $\tri{g}_{ik}=(\dd_i y^a)(\dd_k y_a)$.
At different values of symmetric indices $i,k$ the quantity $\tri{b}^a_{ik}$ defines six vectors in the ambient space, which are orthogonal to the tangent vectors $\dd_i y^a$ according to the properties of the second fundamental form.
In the generic case, there is {the} only vector  $\om^a$ defined by conditions
 \disn{t3}{
\om_a \dd_i y^a=0,\qquad
\om_a \tri{b}^a_{ik}=0,\qquad
|\om^a \om_a|=1.
\nom}
Using it, let us introduce a quantity $\tri{\al}^{ik}_a$, defined by relations close to \eqref{s16} \cite{statja18}:
\disn{t20}{
\tri{\al}^{ik}_a=\tri{\al}^{ki}_a,\qquad
\tri{\al}^{ik}_a \dd_i y^a=0,\qquad
\tri{\al}^{ik}_a \om^a=0,\qquad
\tri{\al}^{ik}_a \tri{b}^a_{lm}=\frac{1}{2}\ls\de^i_l\de^k_m+\de^i_m\de^k_l\rs.
\nom}
Contracting the quantities $\tri{\al}^{ik}_b$ and $\tri{b}^a_{lm}$ the other way, we obtain a projector on a corresponding 6-dimensional space:
\disn{t20.1}{
\tri{\al}^{ik}_b \tri{b}^a_{ik}=\de^a_b-\tri{g}^{ik}(\dd_i y^a)\dd_k y_b-\om^a \om_b \om^c \om_c,
\nom}
where the second and the third terms at the r.h.s. are projectors on 3-dimensional and 1-dimensional spaces, correspondingly.

Using \eqref{n6.12}, let us rewrite \eqref{n6a} in the form
 \disn{nn2}{
(\dd_0 y_a) \tri{b}^a_{ik}=\Ga_{0,ik}-g_{0m}\tri{\Ga}^m_{ik}.
\nom}
Multiplying it by $\tri{\al}^{ik}_b$ and using \eqref{t20.1}, we obtain
 \disn{nn3}{
(\dd_0 y_a)\ls \de^a_b-\tri{g}^{ik}(\dd_i y^a)\dd_k y_b-\om^a \om_b \om^c \om_c\rs=\tri{\al}^{ik}_b\ls\Ga_{0,ik}-g_{0m}\tri{\Ga}^m_{ik}\rs,
\nom}
Using it together with \eqref{n6.12}, we can write
 \disn{nn4}{
\dd_0 y^a=\tri{g}^{ik}g_{0i}\dd_k y^a+\tri{\al}^{aik}\ls\Ga_{0,ik}-g_{0m}\tri{\Ga}^m_{ik}\rs+\dz \om^a,
\nom}
where we denote $\dz=\om^c \om_c\om^a\dd_0 y_a$.
Taking the properties \eqref{t3},\eqref{t20} into account, one can find the quantity $\dz$ by substitution of \eqref{nn4} into \eqref{n6.13}:
 \disn{nn5}{
\dz=\sqrt{\left|
g_{00}-g_{0i}\tri{g}^{ik}g_{0k}-\tri{\al}^{ik}_b\tri{\al}^{bln}\ls\Ga_{0,ik}-g_{0m}\tri{\Ga}^m_{ik}\rs\ls\Ga_{0,ln}-g_{0p}\tri{\Ga}^p_{ln}\rs
\right|}.
\nom}
As a result, we obtain the explicit form of the dynamical equation for the independent variable $y^a$
in the form \eqref{nn4} with \eqref{nn5}.
Together with dynamical equations \eqref{n6.21},\eqref{n6.22} for the independent variables $\ta^{0\n}$
and the constraint \eqref{n6.11} (which could only be imposed on $y^a$
at the initial moment of time $x^0$)
the equation \eqref{nn4} forms a complete set of equations of motion of embedding matter. The equation \eqref{nn4} automatically guarantees the satisfaction of constraint \eqref{n6.11}  at all moments of time after the initial one.

\section{The limiting case of non-relativistic approximation}\label{nonrel1}
Let us analyze the equations of motion of embedding gravity in the \emph{non-relativistic in the bulk} limit \eqref{s7}.
We will assume that  $j^\m_a$ and $\ta^{\m\n}$ (as well as the EMT of matter $T^{\m\n}$) are small enough, so due to the small value of the gravitational constant $\ka$ the r.h.s. of Einstein equations \eqref{r3.1}
is small and gravitational field turns out to be weak:
 \disn{s20}{
g_{\m\n}=\eta_{\m\n}+h_{\m\n},\qquad
h_{\m\n}\to0,
\nom}
where $\eta_{\m\n}$ is a flat metric.
Let us firstly find a solution in a limiting case when $\de j^\m_a=0$ (see \eqref{s7}) and $h_{\m\n}=0$. The corresponding quantities will be marked by a bar.

The limiting value of the set of currents can be written as
 \disn{n7}{
\bar j^\m_a=\de^\np_a \bar\rho_\ta \bar u^\m,
\nom}
where the arbitrary vector $j^\m$ that appeared in  \eqref{s7}   (it must be timelike for the action \eqref{r5} to be real) is presented in the form
 \disn{n8}{
j^\m=\bar\rho_\ta \bar u^\m,\qquad
\bar u^\m \bar u^\n\bar g_{\m\n}=1,
\nom}
where
 \disn{n8.1}{
\bar g_{\m\n}=\eta_{\m\n}.
\nom}
Substituting \eqref{n7} in \eqref{r5.1}, we obtain that
 \disn{n9}{
\bar\ta^{\m\n}=\bar\rho_\ta \bar u^\m \bar u^\n.
\nom}
Substituting \eqref{n7} and \eqref{n9} in \eqref{s6.2}, we notice that
 \disn{n10}{
\de^\np_a \bar\rho_\ta \bar u^\m=\bar\rho_\ta \bar u^\m \bar u^\n\dd_\n \bar y_a\quad\Rightarrow\quad
\bar u^\n\dd_\n \bar y_a=\de^\np_a,
\nom}
where we have assumed that $\bar\rho_\ta\ne0$ (the case $\bar\rho_\ta=0$ corresponds to $\bar\ta^{\m\n}=0$,
i.e. the contribution of embedding matter is absent, so this case is of no interest in the context of the analysis of its properties).

One can always choose coordinates $x^\m$ on a surface defined by embedding function $\bar y^a(x)$ in such a way that $\bar y^0=x^0$.
The relation \eqref{n10} means that the vector $\de^\np_a$ is tangent to the surface at any point of it so that one can impose an additional restriction on a coordinate system:
 \disn{n10.1}{
\dd_0 \bar y^I=0
\nom}
(here and hereafter $I,K,\ldots=1,\ldots,9$). Therefore we can conclude that there is a certain coordinate system such that the embedding function $\bar y^a(x)$ in the considered limit has the form
 \disn{n13}{
\bar y^0=x^0,\qquad \bar y^I=\bar y^I(x^i).
\nom}
On the other side, there is a certain coordinate system (possibly not the same as above, generally speaking)
{in which} the relation \eqref{n8.1} is satisfied, so the curvature tensor of the surface vanishes.
If a 3-dimensional submanifold $x^0=0$ of the surface \eqref{n13} is considered
($\bar y^I(x^i)$ is its embedding function to 9-dimensional space), then, using the Gauss equation
(it connects 3-dimensional components of 4-dimensional curvature tensor with 3-dimensional curvature tensor and the second fundamental form of the considered 3-submanifold of Riemannian 4-space, which is equal to zero in the present case), it can be noticed that its curvature is also equal to zero.
One can therefore transform the metric of the discussed 3-submanifold to the form $\bar g_{ik}=-\de_{ik}$ by choosing the coordinates $x^i$.
{After that the metric of 4-surface becomes to coincide with \eqref{n8.1}, which can be easily seen using \eqref{r1}.}
Therefore there are coordinates in which the relations \eqref{n8.1} and \eqref{n13} are simultaneously satisfied, and we will use such coordinates.

Substituting \eqref{n13} in \eqref{n10} at $a=0$, we obtain that $\bar u^0=1$.
Using this fact in  \eqref{n8} together with \eqref{n8.1}, we find that
 \disn{n11}{
{\bar u^0{}}^2-\bar u^i \bar u^i=1\quad\Rightarrow\quad \bar u^i=0 \quad\Rightarrow\quad \bar u^\m=\de^\m_0.
\nom}
According to \eqref{n9}, this leads to the expression for EMT:
 \disn{s25}{
\bar\ta^{\m\n}=\bar\rho_\ta \de^\m_0 \de^\n_0,
\nom}
which corresponds to the resting dust embedding matter with density $\bar\rho_\ta$.

As it was noticed after  \eqref{n13}, the quantity $\bar y^I(x^i)$ turns out to be an embedding of flat 3-metric in a 9-dimensional Euclidean space. Since generic 3-metric is parametrized by six independent components, such embedding can be parametrized by three arbitrary functions. In a general case (which, for example, has to be realized if we consider the equations of motion of embedding matter as a dynamics w.r.t. time $x^0$ and do not perform a fine-tuning of initial values) this embedding has such form that there is a quantity $\bar\al^{ik}_a$ defined by \eqref{s16} which corresponds to \eqref{n13}, and $\bar\al^{ik}_\np=0$. Such embedding structure corresponds to the notion of \emph{free embedding} introduced in  \cite{bustamante} in context of 3-dimensional manifold, and to the notion of \emph{spatially free embedding} in context of 4-dimensional one. Note that in the considered limit, when the embedding function has the form \eqref{n13}, the quantities defined by relations \eqref{s16} and \eqref{t20} coincide with each other, so $\tri{\bar\al}^{ik}_a=\bar\al^{ik}_a$ and
 \disn{s25.1}{
\bar\om^a=\de^a_0.
\nom}

Using \eqref{n8.1}, \eqref{n13} and \eqref{s25}
in the equations of motion of embedding matter \eqref{n6.11}, \eqref{nn4}, \eqref{n6.21}, \eqref{n6.22}
one can obtain the only restriction of the considered limit:
 \disn{s26}{
\dd_0\bar\rho_\ta=0,
\nom}
which means that the embedding matter density does not change with time. In this case $\bar\rho_\ta$ can arbitrarily depend on spatial coordinates $x^i$; this dependence is determined by the choice of initial values. It can also be noticed that according to  \eqref{nn5} we have $\bar\dz=1$.

\section{Equations of motion in the nonrelativistic limit}\label{nonrel2}
Now let us analyze the equations of motion in such regime that the independent variables are close to their limiting values considered in the previous Section. In this case the metric has the form \eqref{s20}. We will additionally assume that ordinary matter is a dust with density $\rho$ and non-relativistic motion, so its EMT in the leading order has the same form as \eqref{s25}:
 \disn{u00}{
T^{\m\n}=\rho \de^\m_0 \de^\n_0,
\nom}
and $\dd_0\rho$ is small. Taking into account that EMT of embedding matter $\ta^{\m\n}$ is close to \eqref{s25} and using harmonic coordinates, as it is usually done in such cases, one can write down linearize Einstein equations \eqref{r3.1} as follows:
 \disn{u00.1}{
\square h_{\m\n}=-\ka(\rho+\ta^{00})\de_{\m\n}.
\nom}
It leads to the fact that in the leading order $h_{\m\n}$ is diagonal and their components are proportional to $\ka$, whereas its non-diagonal elements are of higher order, so, in particular,  $h_{0i}$ is small in comparison with $\ka$.
Note also that the small value of $\dd_0\rho$ together with \eqref{s26} leads to the slow change of gravitational field, so $\dd_0 h_{\m\n}$ is also small in comparison with $\ka$.

If we consider a finite time interval $\De x^0$ only, then the embedding function  $y^a$ can be obtained as a small deviation from its limiting value \eqref{n13}:
 \disn{u0}{
y^a=\bar y^a+\de y^a,
\nom}
where $\de y^a$ is small.
Note that the expression \eqref{n13} gives an explicit expression of the component $y^0$ only, while for remaining components, it merely means that they are time-independent according to the condition \eqref{n10.1}.
In the case that is close to the limiting one, this condition is replaced by the smallness of $\dd_0 y^I$. Therefore, if the behavior of independent variables is studied not only at finite, but also at large time intervals; in particular, at those intervals, $\De x^0$ whose inverse is proportional to the smallness parameter (we will do it below in the transition to the non-relativistic limit), then $y^I$ could start to change noticeably, and it turns out to be impossible to write it down as \eqref{u0}  with a small
$\de y^I$.
For that reason, we will assume that the only component that can be written as its limiting value plus a small deviation is $y^0$, and the remaining components are restricted only by the condition that their derivatives w.r.t. $x^0$ are small:
 \disn{u1}{
y^0=x^0+w,\quad y^I=y^I(x^0,x^i);\qquad
\dd_0 y^I(x^0,x^i)\ll 1,
\nom}
where $w$ is small.
For finite intervals $\De x^0$, such formulation of conditions corresponds to the expression \eqref{u0}
with small $\de y^a$ and, at the same time, allows components $y^I$ to change noticeably at the large (i.e., of the order of inverse values of $\dd_0 y^I$ components) intervals $\De x^0$.
Note that we do not choose any particular coordinate system when writing the embedding function in the form \eqref{u1}, so the metric
$g_{\m\n}$ can be defined in any coordinates. This fact was used above \eqref{u00.1} when harmonic coordinates had been chosen.

In the considered equations of motion of embedding matter  (namely, in \eqref{n6.22} and \eqref{nn4}) the quantities $\al^{ik}_a$ and $\tri{\al}^{ik}$ are present, which are defined by relations \eqref{s16} and \eqref{t20}.
As was noted above \eqref{s25.1}, in the limiting case
{(and hence in the leading approximation, when we consider the deviations from the limiting case)}
these quantities are equal to each other and vanish at $a=0$.
We will denote their values in the leading approximations as $\hat\al^{ik}_a$, using the following exact definition of it: \disn{u2}{
\hat\al^{ik}_a=\hat\al^{ki}_a,\qquad
\hat\al^{ik}_\np=0,\qquad
\hat\al^{ik}_I \dd_m y^I=0,\qquad
\hat\al^{ik}_I \dd_l\dd_m y^I=\frac{1}{2}\ls \de^i_l\de^k_m+\de^i_m\de^k_l\rs,
\nom}
which is the limit of \eqref{s16} and \eqref{t20}. Defined as such, the leading approximation of the quantities $\al^{ik}_a$ and $\tri{\al}^{ik}_a$ coincides with time-independent quantity $\bar\al^{ik}_a$ for the embedding function \eqref{n13}. However, after its replacement  by  \eqref{u1} this coincidence is no longer exact due to the appearance of a slight dependence of $y^I$ on $x^0$.
As a result, $\hat\al^{ik}_a$ can also be slightly time-dependent (by slightness we mean that $\dd_0\hat\al^{ik}_a$ is small due to the satisfaction of the condition from \eqref{u1}).

Let us start the analysis of the equations of motion of embedding matter from the equation \eqref{nn4} which defined a dynamics of embedding function.
To do that, let us write down the formula \eqref{t20.1} at $a=0$ and $b=I$ using \eqref{nn1}:
 \disn{u3}{
\tri{\al}^{ik}_I \ls \dd_i\dd_k y^0-\tri{\Ga}^m_{ik}\dd_m y^0\rs=
-\tri{g}^{ik}(\dd_i y^0)\dd_k y_I-\om^0 \om_I,
\nom}
where we took into account that
 \disn{u3.1}{
\om^a \om_a=1
\nom}
according to  \eqref{t3},\eqref{s25.1} and the small value of deviations.
It follows from \eqref{u3} that with the use of  \eqref{s20}, \eqref{s25.1} and \eqref{u1}, in the leading approximation one can find:
 \disn{u4}{
\om^I=(\dd_i w)\dd_i y^I+\hat{\al}^{ik}_I \dd_i\dd_k w.
\nom}
Now let us find the leading approximation of \eqref{nn4}, assuming that $\ka$, as well as $w$, is small, and we can neglect higher powers of them. Using \eqref{nn5}, \eqref{s20}, \eqref{u1} and \eqref{u3.1}
together with the fact that $h_{0i}$ and $\dd_0 h_{\m\n}$ are small in comparison with $\ka$ (see below \eqref{u00.1}),
in such approximation we obtain
 \disn{u5.1}{
\dd_0 w=\frac{1}{2}\ls h_{00}+\om^I\om^I\rs,
\nom}\vskip -2em
 \disn{u5.2}{
\dd_0 y^I=\om^I+o(\ka),
\nom}
where the quantity $\om_I$ is given bt \eqref{u4} and has the same order of magnitude as $w$.

Writing down the remaining dynamical equations \eqref{n6.21}, \eqref{n6.22}, at the leading order in a similar manner, we obtain
 \disn{u6.1}{
\dd_0 \rho_\ta=-\dd_i \ta^{0i}+o(\ka),
\nom}\vskip -2em
 \disn{u6.2}{
\dd_0 \ta^{0m}
=-\rho_\ta\Ga^m_{00}+
\dd_l \Biggl(\hat\al^{lm}_I \biggl(\rho_\ta\hat\al^{ik}_I\Bigl((\dd_i \om_L)(\dd_k \om_L)-R_{0i0k}\Bigr)+2\ta^{0i}\dd_i \om^I\biggr)\Biggr),
\nom}
where we denote $\rho_\ta\equiv\ta^{00}$. We used
 \eqref{u1},\eqref{u2},\eqref{u5.1},\eqref{u5.2} and took into account that
 $h_{0i},\dd_0 h_{\m\n}\sim o(\ka)$.

Let us examine the time evolution of independent variables governed by equations \eqref{u5.1}-\eqref{u6.2}.
Consider the situation when at the initial moment the variables $w$ a $\ta^{0i}$
(as well as $\om^I$ according to \eqref{u4}) are either vanished or very small.
While this condition is satisfied well enough, the contributions at the r.h.s. of \eqref{u5.1} and \eqref{u6.2} containing these quantities can be neglected in comparison with remaining contributions of order $\ka$. The time dependence of the latter can also be neglected, so with these assumptions, the r.h.s. of \eqref{u5.1} and \eqref{u6.2} will also be time-independent in the leading order, and $w$, as well as $\ta^{0i}$, will increase linearly w.r.t. $x^0$. In a finite time interval $\De x^0$ they will necessarily reach the order of magnitude of $\ka$,
{but despite this}
the contributions in r.h.s of \eqref{u5.1}, \eqref{u6.2}, in which they are present, will be nevertheless small in comparison with leading contributions of order $\ka$, so the linear increasing of $w$ and $\ta^{0i}$ will be continued. Consequently, according to \eqref{u5.2} and\eqref{u6.1}, the quantities $y^I$ and $\rho_\ta$ will increase quadratically with time $x^0$, i.e. will receive an additive perturbation proportional to ${x^0}^2$.

Such \emph{trivial} (i.e. corresponding to a linear or quadratic time dependence) dynamics would take place during an arbitrary \emph{finite} interval of time $\De x^0$, while the quantities $w$ and $\ta^{0i}$ have order of $\ka$. However, for a time $\De x^0\sim 1/\sqrt{\ka}$ (which is not assumed to be finite anymore, since $\ka$ is small), these quantities would increase to the order
$\sqrt{\ka}$. After that, it would become impossible to drop out the contributions at the r.h.s. of equations \eqref{u5.1} and \eqref{u6.2}, which were neglected earlier. Note also that after non-finite time intervals $\De x^0$,
{it is no longer possible to}
assume time independence of the contributions
{to the r.h.s. of \eqref{u5.1} and \eqref{u6.2},}
that were initially of the order of $ \ka $.

Therefore the dynamics defined by equations \eqref{u5.1}-\eqref{u6.2} becomes nontrivial at time scales of order $\De x^0\sim 1/\sqrt{\ka}$.
In order to analyze this nontrivial dynamics (which can easily be recognized as a standard non-relativistic one) let us perform a time change, introducing a large quantity $c$:
 \disn{s51}{
x^0=c\,t,\qquad
c\sim  \frac{1}{\sqrt{\ka}}.
\nom}
One can assume that $c$ is the speed of light, then, taking into account that
 \disn{s52}{
\ka=\frac{8\pi G}{c^2},
\nom}
it must be assumed that Newtonian gravitational constant  $G$ is a finite quantity, whereas the small value of $\ka$ corresponds to the large value of  $c$ according to \eqref{s52}.

Using the fact that $\dd_0 h_{\m\n}$ is small in comparison with  $\ka$ (see below \eqref{u00.1}), one can obtain from \eqref{u00.1} that at the leading order
 \disn{s52.1}{
h_{\m\n}=\frac{2\ff}{c^2}\de_{\m\n},
\nom}
where $\ff$ is the Newtonian potential corresponding to matter distribution with density $\rho+\rho_\ta$:
 \disn{s53a4}{
\Delta\ff=4\pi G (\rho+\rho_\ta).
\nom}
As it was mentioned above, in the nontrivial regime the independent variables $w$ and $\ta^{0i}$ have order of $\sqrt{\ka}\sim 1/c$. Due to this, in order to analyze non-relativistic limit, let us introduce some new quantities instead of them:
 \disn{u7}{
\ps=cw,\qquad
v_\ta^i=c\frac{\ta^{0i}}{\rho_\ta},
\nom}
which remain finite at $\ka\to0$, $c\to\infty$.
Let us also introduce
 \disn{u8}{
\ga^I=c\om^I=(\dd_i \ps)\dd_i y^I+\hat{\al}^{ik}_I \dd_i\dd_k \ps,
\nom}
instead of $\om^I$, where \eqref{u4} was used.
In terms of these new variables  and derivatives w.r.t. new \emph{non-relativistic time} $t$
 \disn{s54}{
\dd_t\equiv \frac{\dd}{\dd t}=c\dd_0
\nom}
the equations \eqref{u5.1}-\eqref{u6.2} can be rewritten up to terms of order $1/c$ as follows:
 \disn{u9.1}{
\dd_t \ps=\ff+\frac{1}{2}\ga^I\ga^I,
\nom}\vskip -2em
 \disn{u9.2}{
\dd_t y^I=\ga^I,
\nom}\vskip -2em
 \disn{u9.3}{
\dd_t \rho_\ta=-\dd_i (\rho_\ta v_\ta^i),
\nom}\vskip -2em
 \disn{u9.4}{
\dd_t (\rho_\ta v_\ta^m)=-\rho_\ta \dd_m\ff+
\dd_l \Biggl(\rho_\ta\hat\al^{lm}_I \biggl(\hat\al^{ik}_I\Bigl((\dd_i \ga^L)(\dd_k \ga^L)+\dd_i\dd_k\ff\Bigr)+2v_\ta^i\dd_i \ga^I\biggr)\Biggr),
\nom}
where we used \eqref{s52.1}, \eqref{u7} and \eqref{u8}.

The resulting dynamical equations \eqref{u9.1}-\eqref{u9.4} describe the motion of the embedding matter in the non-relativistic limit $c\to\infty$, and, together with \eqref{s52.1},\eqref{s53a4} (to which the Einstein equations are reduced in the considered case)
{and the equations of motion of ordinary matter}
fully describe the non-relativistic dynamic of a system. The possibility to construct consistent non-relativistic equations means, in particular, that non-relativistic approximation, appearing as a result of non-relativistic character of currents \eqref{s7} in the ambient space, turns out to be stable: the embedding matter remains non-relativistic.

The dynamic equations \eqref{u9.1}-\eqref{u9.4} need to be supplemented by imposition of the constraint \eqref{n6.11} at the initial moment, which (if \eqref{s20},\eqref{u1},\eqref{u7} are taken into account), up to terms of order $1/c$, takes the form
 \disn{u10}{
(\dd_i y^I)(\dd_k y^I)=\de_{ik}.
\nom}
This equation means that  in the non-relativistic limit   (we remind that the time conservation of the constraint is provided by dynamical equations, see below \eqref{n6a})
$y^I$ resembles an embedding function of flat 3-dimensional metric in 9-dimensional ambient space at any moment of time.
As we mentioned above (see the text below \eqref{s25}), such embedding is parametrized by three arbitrary functions. Unfortunately, we did not succeed in the construction of an explicit form of such functions. These functions evolve with time according to the dynamical equation
\eqref{u9.2}, which can be interpreted as dynamical isometric bending of a 3-dimensional surface with a flat Euclidean metric. It can be straightforwardly checked that the deformation governed by  \eqref{u9.2} is indeed isometric bending.

We obtain that the embedding matter in the non-relativistic limit is described by the following eight parameters:
\begin{itemize}
	\item three functions which parametrize an embedding function {of a flat 3-dimensional Euclidean metric in a 9-dimensional Euclidean space,}
	\item the quantity $\ps$  (see \eqref{u7} and \eqref{u1}),
	\item the density $\rho_\ta$ of the embedding matter,
	\item the velocity $v_\ta^i$ of the embedding matter.
\end{itemize}
To obtain an equation that describes the velocity $v_\ta^i$ dynamics, \eqref{u9.4} needs to be rewritten using \eqref{u9.3}. Let us do that, together with the formulation of this equation in a familiar way of dust matter description, where it is not just $\dd_t v_\ta^m$ that is present at the l.h.s. of the equation, but rather some combination resembling an acceleration of a "single particle"{} of the embedding matter (although we assume that in the fundamental theory such particles do not exist, and all variables describing the embedding matter are, in fact, gravitational variables in the framework of embedding gravity).
As a result, we obtain:
 \disn{u11}{
\rho_\ta (\dd_t  + v_\ta^i\dd_i ) v_\ta^m=
-\rho_\ta \dd_m\ff+
\dd_l \Biggl(\rho_\ta\bigg[v_\ta^l v_\ta^m+\hat\al^{lm}_I \biggl(\hat\al^{ik}_I\Bigl((\dd_i \ga^L)(\dd_k \ga^L)+\dd_i\dd_k\ff\Bigr)+2v_\ta^i\dd_i \ga^I\biggr)\bigg]\Biggr).
\nom}

The non-relativistic equations of motion of the embedding matter in the form \eqref{u9.1}-\eqref{u9.3},\eqref{u11} reproduce the ones that were obtained by the more complicated (and less geometrically clear) way in the work \cite{statja67}.
The equation \eqref{u9.3} is a usual conservation law of this matter, and
\eqref{u11} defines the force (per unit volume) acting on its "single particles"{}. Its first term is a well-known gravitational force corresponding to Newtonian approximation. The remaining terms can be interpreted as a certain self-interaction force of the embedding matter, which depend not only on familiar characteristics of this matter, namely its density $\rho_\ta$ and velocity $v^i_\ta$, but also on its additional characteristics $\ps$ and $y^I$,
whose dynamics is governed by equations \eqref{u9.1} and \eqref{u9.2}.

\section{Embedding function in the non-relativistic limit}\label{geom}
Let us find the approximate expression for the embedding function $y^a(x)$ corresponding to the
non-relativistic limit
that was found in the previous Section. According to  \eqref{u9.1} and \eqref{u9.2}, the quantities $\ps$ and $y^I$ have certain dependence on $t=x^0/c$ governed by these equations.
Therefore, according to \eqref{u1} together with \eqref{u7}, the components of the embedding functions can be written as
 \disn{u12}{
y^0=x^0+\frac{1}{c}\,\ps\ls\frac{x^0}{c},x^i\rs,\qquad
y^I=y^I\ls\frac{x^0}{c},x^i\rs
\nom}
(note that in this notation the condition \eqref{u1} is satisfied),
where the quantities $\ps(t,x^i)$, $y^I(t,x^i)$
satisfy the equations that are reduced to \eqref{u9.1}, \eqref{u9.2} and \eqref{u10} in the limit $c\to\infty$.
The limiting forms {of these quantities} that \emph{exactly} satisfy the mentioned equations,
are denoted as $\hat\ps(t,x^i)$, $\hat y^I(t,x^i)$  in this Section,
so one can write
 \disn{u13.1}{
\ps(t,x^i)=\hat\ps(t,x^i)+\de\ps(t,x^i),
\nom}\vskip -2em
 \disn{u13.2}{
y^I(t,x^i)=\hat y^I(t,x^i)+\de y^I(t,x^i),
\nom}
where $\de\ps(t,x^i)$ and $\de y^I(t,x^i)$ are small corrections.
According to \eqref{u10},  $\hat y^I(t,x^i)$ at each value of $t$ is an embedding function of a flat 3-dimensional Euclidean metric in 9-dimensional Euclidean space, so the corresponding induced metric exactly coincides with $\de_{ik}$.
Note that an arbitrary perturbation of this induced metric can be obtained by choosing the embedding function deformation in the form
 \disn{u14}{
\de y^I(t,x^i)=\xi_{ik}\hat \al^{Iik}.
\nom}
Indeed, taking \eqref{u2} into account, in the leading approximation  we obtain
 \disn{u15}{
\de \ls(\dd_i y^I)(\dd_k y^I)\rs=\ls\dd_l \hat y^I\rs\dd_m \ls\xi_{ik}\hat \al^{Iik}\rs+\ls\dd_m \hat y^I\rs\dd_l \ls\xi_{ik}\hat \al^{Iik}\rs+o(\xi)=2\xi_{lm}+o(\xi).
\nom}

Let us substitute the embedding function \eqref{u12} (using \eqref{u13.1} and \eqref{u13.2}) into the induced metric formula \eqref{r1}.
In all calculations we will omit terms which are small in comparison with $1/c^2$, and we assume that up to this order $\de\ps(t,x^i)=0$ and $\de y^I(t,x^i)$ reduces to \eqref{u14}.
For the $\{ik\}$ components of the induced metric we, using \eqref{u15}, obtain
 \disn{s69.3}{
g_{ik}=\frac{1}{c^2}(\dd_i \ps)(\dd_k \ps)-(\dd_i y^I)(\dd_k y^I)
=\frac{1}{c^2}(\dd_i \hat\ps)(\dd_k \hat\ps)-\de_{ik}-2\xi_{ik}+o(\xi).
\nom}
To match this quantity with \eqref{s20} (taking \eqref{s52.1} into account) up to the desired order, i.e. with $-\de_{ik}+2\ff\de_{ik}/c^2$,
one must take
 \disn{u16}{
\xi_{ik}=\frac{1}{c^2}\ls\frac{1}{2}(\dd_i \hat\ps)(\dd_k \hat\ps)-\ff\de_{ik}\rs.
\nom}
For the $\{0k\}$ components of the induced metric, using  \eqref{u9.2} and \eqref{u8}, we obtain
 \disn{s69.2}{
g_{0k}=\ls 1+\frac{1}{c^2}\, \dd_t\ps\rs\frac{1}{c}\dd_k \ps-\frac{1}{c}(\dd_t y^I)(\dd_k y^I)=
\frac{1}{c}\dd_k \hat\ps-\frac{1}{c}(\dd_t \hat y^I)(\dd_k \hat y^I)=0,
\nom}
which corresponds to \eqref{s52.1} with the above precision.
Lastly, for the $\{00\}$ component, using \eqref{u9.1},\eqref{u9.2}, we find
 \disn{s69.1}{
g_{00}=\ls 1+\frac{1}{c^2}\, \dd_t\ps\rs^2-\frac{1}{c^2} (\dd_t y^I) (\dd_t y^I)=
1+\frac{2}{c^2}\dd_t\hat\ps-\frac{1}{c^2} (\dd_t\hat y^I) (\dd_t\hat y^I)=
1+\frac{2\ff}{c^2},
\nom}
which also corresponds to \eqref{s52.1} with the above precision.

Since it is always possible to make any metric perturbation of the order higher than $1/c^2$ by addition of such terms to the embedding function, one can conclude that the embedding function
 \disn{s68}{
y^0=x^0+\frac{1}{c}\,\hat\ps\ls\frac{x^0}{c},x^i\rs+o\ls\frac{1}{c^2}\rs,\no
y^I=\hat y^I\ls\frac{x^0}{c},x^i\rs+\frac{1}{c^2}\hat\al^{Iik}\ls\frac{1}{2}(\dd_i \hat\ps)(\dd_k \hat\ps)-\ff\de_{ik}\rs+o\ls\frac{1}{c^2}\rs
\nom}
corresponds to the metric of the non-relativistic limit of embedding gravity obtained in the previous Section.
The approximate expression \eqref{s68} for embedding function reproduces the one that was found in the work \cite{statja67}.

\section{Conclusion}\label{zakl}
Let us briefly summarize the results we have obtained.
We study the possibility to explain the mystery of DM through the transition from GR to embedding gravity, which is a modified gravity based on a simple string-inspired geometric method. After the reformulation of this theory in the form of GR with an additional contribution of the embedding matter, we find its equations of motion in the general case.
These equations are reduced to a set of first-order dynamical equations \eqref{n6.21},\eqref{n6.22},\eqref{nn4} and the constraint \eqref{n6.11} that is in involution with them.
One can try to explain various effects in terms of embedding matter by analyzing these equations in different regimes. Such effects could be related not only to DM (which is the topic of the present work) but also {probably} to the dark energy and the inflation.

Since observational discrepancies, which exist in the framework of GR, can be quite well explained by an assumption that DM is a cold dust matter, we consider the class of solutions of embedding gravity, which corresponds to the non-relativistic character of motion of the embedding matter. This class of solutions naturally appears due to the assumption that conserved currents $j_\m^a$ describing the embedding matter are non-relativistic in the bulk \eqref{s7}. The obtained non-relativistic regime of motion of fictitious embedding matter turns out to be stable in such sense that it remains non-relativistic during the evolution if the choice of initial values corresponds to the non-relativistic regime.
Especially interesting is the fact that in the non-relativistic regime, besides usual coupling with gravity, this embedding matter possesses certain self-interaction, as in the models of Self-Interacting Dark Matter \cite{1705.02358}, which could be perspective in the context of solving the core-cusp problem
and other observational problems on the scales of galaxies appearing in the $\La$CDM model.

In the present work, we use the method of obtaining non-relativistic equations of motion, which has a simpler geometric meaning than the one that was used in the work \cite{statja67},
since from the start to the end of the calculations, we use 4-dimensional embedding function  $y^a(x^\m)$.
It should be noted that in the present work, as well as in \cite{statja67}, the choice of the spatial part of an embedding function
is essentially different from the one that was used at the cosmological scales in the works \cite{davids01,statja26}, where it has minimal codimension (i.e., the 3-dimensional surface was embedded in 4-dimensional flat bulk). On the contrary, in the present work, we suppose that for a random choice of initial values, this 3-dimensional surface turns out to be "unfolded"{} in 9-dimensional flat bulk, which corresponds to the existence of the quantity $\hat \al^{ik}_I$ defined in \eqref{u2}, so the 4-dimensional embedding is
\emph{spatially free} \cite{bustamante}.

In the non-relativistic limit, the embedding matter is described by eight fields: besides its density $\rho_\ta$ and velocity $v^i_\ta$ an additional field $\ps$ is present alongside with an embedding function $y^I$ of a flat 3-dimensional {Euclidean} metric in 9-dimensional {Euclidean} space, which form is described by three functions. These four additional degrees of freedom have their own dynamics and affect the process of self-interaction of the embedding matter. An important role in the embedding matter's behavior is played by the initial values of these variables, which correspond to their values at the beginning of the non-relativistic regime. These values are formed during the preceding relativistic period, which possibly corresponds to the inflationary era.

{\bf Acknowledgements.}
The author is grateful to A.~Sheykin for useful discussion.
The work is supported by RFBR Grant No.~20-01-00081.


\end{document}